# Wigner Crystallization of Rotating Dipolar Fermions in the Fractional Quantum Hall Regime

Szu-Cheng Cheng,[1,*] Shih-Da Jheng[2] and T. F. Jiang[2]

[1]*Department of Physics, Chinese Culture University, Taipei 11114, Taiwan*

[2]*Institute of Physics, National Chiao Tung University, Hsinchu 30010, Taiwan*

Abstract

We show the possible existence of the Wigner crystal (WC) in the Fractional Quantum Hall (FQH) regime. We find that the Landau-level mixing (LLM) will lower the energy of the WC significantly in the high-density regime. The WC is lower in energy than the FQH liquid in the high-density regime. We conclude that the crystal phase is expected at *high density* for rotating dipolar gases, which is consistent with non-rotating dipolar gases, but is inconsistent with the *low-density* conclusion from Baranov *et al.* [Phys. Rev. Lett. **100**, 200402 (2008)], where the effect of LLM is ignored.

[*] Electronic address: sccheng@faculty.pccu.edu.tw; Fax: +886-2-28610577

The recent success in the creation of chromium Bose-Einstein condensate (BEC) [1, 2] and progress in realization heteronuclear polar molecules [3-8] have attracted growing interest in quantum degenerate dipolar gases. The anisotropic character and long-rang nature of the dipole-dipole interactions (DDI) makes the dipolar systems different from the systems described by contact interactions [9, 10]. For the dipolar BEC, new quantum phases are predicted [11, 12]. The influence of the trapping geometry on the stability of the BEC and the effect of the DDI on the excitation spectrum are investigated [13]. The vortex lattice of rotating dipolar BEC exhibits novel bubble, stripe, and square structures [14]. For the Fermi gases, the $s$-wave scattering is prohibited due to the Pauli Exclusion Principle. This principle also implies a strong suppression of three-body loss [15], and consequently the strong interactions of fermions are achieved by the Feshbach resonance where the scattering length takes a divergent value [16, 17]. Bond pairs of fermions with resonant interaction are formed and the system of a Fermi gas behaves as a bosonic gas of molecules [18, 19]. The observed pairing of fermions provides the crossover between the weakly-paired, strongly overlapping Bardeen-Cooper-Schrieffer regime, and the tightly bound, weakly-interacting diatomic molecular BEC regime [20]. The strong correlations of fermions induced by the dipolar interaction can then be explored [21]. There is dipolar-induced superfluidity [22, 23] and fractional quantum Hall (FQH) states in rotating dipolar Fermi gases [24, 25].

In a system with interaction potential energy dominating the kinetic energy, the system will adopt a configuration of the lowest potential energy and crystallize into a lattice that is termed the Wigner crystal (WC) [26]. For non-rotating dipolar gases, the crystal phase is expected at high density [11, 12]. Rotating fermions with the inertial mass $M$ and the rotational frequency $\omega$ feel the Coriolis force in the rotating frame. The Coriolis force on rotating gases is equivalent to the Lorentz force of a charged particle in a magnetic field. Quantum-mechanically, energy levels of a charged particle in a uniform magnetic field show discrete Landau levels. In the lowest Landau level, the kinetic energy of rotating dipolar gases is frozen and the DDI create strong correlations on particles. Therefore, in the lowest

Landau level, the potential energy from DDI dominates the kinetic energy and rotating dipolar gases will crystallize into a WC. But it has demonstrated that the WC is stable in the $v<1/7$ or low density regime [25], if only the lowest Landau level and no Landau-level mixing (LLM) are considered. Here the filling factor $v \equiv 2\pi\rho\lambda^2$, which measures the fraction of Landau levels being occupied by particles, where $\rho$ and $\lambda = \sqrt{\hbar/2M\omega}$ are the average density and the magnetic length, respectively.

In fact, in the case of an electron gas, a study of Wigner crystallization has confirmed that the reentrant behavior to an insulating phase in the FQH regime is a consequence of an increased stability of the WC due to the effects from LLM [27]. The lowering in the WC energy from LLM is substantially larger than that in the FQH liquid. The amount of LLM is affected by the two relevant energies, which are the inter-Landau-level spacing $\hbar\omega_c$ and the dipolar interaction $E_d = D/a^3$, where $\omega_c = 2\omega$ is the cyclotron frequency, $D$ is a measure of the strength of DDI and $a = 1/\sqrt{\rho}$ is the average distance between fermions. In effective atomic units the dimensionless length $r_s = a_B/a$, where the Bohr radius $a_B = MD/\hbar^2$. The strength of LLM is related to the LLM parameter $\gamma = E_d/\hbar\omega_c = vr_s/2\pi$. $r_s$ is equivalent to the LLM parameter $\gamma$ as $v$ is fixed. There is a strong LLM in the FQH regime if $r_s$ is big. The large $r_s$ value corresponds to the high density regime, larger particle mass and strong DDI. Therefore, without considering LLM, we still cannot conclude that the WC of rotating dipolar gases is stable in the low-density regime [25]. We shall consider the effect from LLM in studying the phase transition between the FQH liquid and the WC.

In this paper we study the two-dimensional (2D) WC of rotating dipolar fermions in the FQH regime. By constructing a variational wave function for the WC of the rotating 2D dipolar fermions, the effect of LLM on the total energy and its dependence on the density and thickness of the sample are calculated. Theoretical estimates of the critical density for the liquid-solid phase transition are obtained by comparing the energies of the FQH liquid and the WC. We find that the Landau-level mixing will lower the energy of the WC significantly in the high-density regime. The WC has a lower/higher energy than the quantum Hall liquid in the high-density/low-density regime, respectively.

We now describe our variational calculations of the total energy of the WC. We consider polarized dipolar fermions in a rotating cylindrical trap and polarized along the negative $\mathbf{e}_z$ direction of rotation, where $\mathbf{e}_z$ is the unit vector of the $z$-axis. For the sake of simplicity, we will consider a pancake system such that the motion of rotating fermionic gases along the $z$-axis is frozen to the ground state of the axial harmonic oscillator. The sample thickness or the extension of dipolar gases in the axial direction is $d = \sqrt{\hbar/M\omega_z}$, where $\omega_z$ is the trapping frequency along the $z$-axis. The Hamiltonian of a quasi-2D rotating dipolar Fermi gas is $H = \sum_i (1/2M)\left[-i\hbar\nabla_i - \mathbf{A}_i\right]^2 + V_D$, where $\mathbf{A}_i = M\omega\,\mathbf{r}_i \times \mathbf{e}_z$ and $V_D = \sum_{i<j} U(\mathbf{r}_i - \mathbf{r}_j)$ describes the effective DDI in 2D. The DDI potential $U(\mathbf{r}) = (D/L^2\lambda^3)\sum_{\mathbf{q}} U(\mathbf{q})\exp(i\mathbf{q}\cdot\mathbf{r})$, where $\mathbf{q}$ is the wave vector and $L$ is the length of the system. $U(\mathbf{q}) = (4\sqrt{2\pi}\lambda/3d) - 2\pi q\exp(\xi^2)\operatorname{Efrc}(\xi)$ is the Fourier transform of the quasi-2D dipolar-interaction potential [13], where $\xi = qd/\sqrt{2}\lambda$ and $\operatorname{Erfc}(\xi) = 1 - \left(2/\sqrt{\pi}\right)\int_0^\xi \exp(-t^2)dt$.

The variational wave function of a particle localized around $\mathbf{r} = \mathbf{R}$ is given by

$$\langle \mathbf{r}|\mathbf{R}\rangle = \sqrt{\frac{2\alpha}{\lambda^2\pi}}\exp\left[-\frac{\alpha}{\lambda^2}(\mathbf{r}-\mathbf{R})^2 + \frac{1}{2\lambda^2}i(\mathbf{r}\times\mathbf{R})\cdot\mathbf{e}_z\right], \qquad (1)$$

where $\alpha$ is a variational parameter to be optimized. The variational wave function is the eigenstate, with eigenvalue $\hbar\omega_c/2$, of the kinetic energy operator as $\alpha=1/4$. This wave function was used by Zhu and Louie in a variational quantum Monte Carlo study of the effect of LLM on stabilizing Wigner crystallization [27]. Our ansatz wave function for the ground state of the WC is a Slater determinant constructed by the wave function of Eq. (1) located at a regular 2D lattice. We have $\Psi(\{\mathbf{r}_i\}) = \det\left|\langle \mathbf{r}_i | \mathbf{R}_j \rangle\right|/\sqrt{N!}$, where $N$ is the number of particles. Similar to the classical dipole lattice, our 2D lattice has a triangular structure with particles centered at sites $\mathbf{R}_j = j_1 \mathbf{a}_1 + j_2 \mathbf{a}_2$, where $\mathbf{a}_1 = a_c(1,0)$ and $\mathbf{a}_2 = a_c\left(1, \sqrt{3}\right)/2$ are the primitive translation vectors of the lattice; and $j_1$ and $j_2$, are any integers to which we refer collectively as $j$. Here $a_c$ is the lattice constant. The area $A_c$ of the primitive unit cell of the triangular lattice is $\sqrt{3}a_c^2/2$.

The overlapping integral $\langle \mathbf{R}_i | \mathbf{R}_j \rangle$ of variational wave functions between two different sites of the lattice is non-zero. We cannot use the Hartree-Fock equation for orthonormal wave functions to calculate the ground-state energy. The total energy per particle, $E_c$, of the WC can be expressed in terms of elements $T_{ij}$ of matrix $\mathbf{T}$, which is the inverse matrix of the matrix with elements $\langle \mathbf{R}_i | \mathbf{R}_j \rangle$ [28]. We obtain $E_c = K_c + V_c(\nu)$, where $K_c$ is the kinetic energy per particle and, as a function of $\nu$, $V_c(\nu)$ is the DDI energy per particle, respectively; and they are given as

$$K_c = \frac{1}{N} \sum_{ij} T_{ji} \left\langle \mathbf{R}_i \left| \frac{1}{2M}[-i\hbar \nabla - M\omega\, \mathbf{r} \times \mathbf{e}_z]^2 \right| \mathbf{R}_j \right\rangle, \qquad (2)$$

$$V_c(\nu) = \frac{1}{2N} \sum_{ijnk} \left(T_{ji}T_{kn} - T_{jn}T_{ki}\right) \langle \mathbf{R}_i, \mathbf{R}_n | U(\mathbf{r}_1 - \mathbf{r}_2) | \mathbf{R}_j, \mathbf{R}_k \rangle. \qquad (3)$$

Equations (2) and (3) are Löwdin's formula [28] for the non-orthogonal wave functions. If wave functions are orthonormal, then $T_{ij} = \delta_{ij}$. The formula $V_c(\nu)$ is just a well-known formula in the Hartree-Slater-Fock theory. From the translational invariant properties of Löwdin's formula [29], we have applied $T_{ij} = T_{(i-j)0} \exp\left(i\, \mathbf{R}_i \times \mathbf{R}_j \cdot \mathbf{e}_z / 2\lambda^2\right)$ and the transformation

$\Sigma_j \exp(i\mathbf{q}\cdot\mathbf{R}_j)/L^2 = \Sigma_\mathbf{G} \delta_{\mathbf{G},\mathbf{q}}/A_c$ to equations (2) and (3), where the **G**'s are the reciprocal lattice vectors. Equations (2) and (3) become:

$$K_c = \sum_j T_{j0} \left\langle 0 \left| \frac{1}{2M}[-i\hbar\nabla - M\omega\,\mathbf{r}\times\mathbf{e}_z]^2 \right| \mathbf{R}_j \right\rangle, \qquad (4)$$

$$V_c(\nu) = \frac{D}{2A_c\lambda^3} \sum_{jk,\mathbf{G}} U(\mathbf{G}) T_{j0} T_{k0} \left\langle 0,0 \left| e^{i\mathbf{G}\cdot(\mathbf{r}_1-\mathbf{r}_2)} \right| \mathbf{R}_j, \mathbf{R}_k \right\rangle$$

$$-\frac{1}{2}\sum_{ijk} T_{j0}T_{k0} \left\langle 0,\mathbf{R}_i \left| U(\mathbf{r}_1-\mathbf{r}_2) \right| \mathbf{R}_j+\mathbf{R}_i,\mathbf{R}_k \right\rangle \exp\left(\frac{-i}{2\lambda^2}\mathbf{R}_i\times\mathbf{R}_j\cdot\mathbf{e}_z\right). \qquad (5)$$

Finite overlapping integrals of variational wave functions between lattice sites are then used to calculate the **T** matrix numerically and are accurate enough to simulate an infinite lattice site.

In calculating the total energy of the WC, we chose 61 lattice sites around the origin to calculate the **T** matrix numerically. Using formulas for $K_c$ and $V_c(\nu)$ at fixed dimensionless length $r_s$ and sample thickness $d$, we can calculate the total energy per particle as a function of variational parameter $\alpha$ in the FQH regime. To test our calculations we can do three things. Firstly, we obtain the kinetic energy $K_c$ being minimum and equal to $\hbar\omega_c/2$ as $\alpha=1/4$. This is consistent with the variational principle that the kinetic energy with $\alpha \neq 1/4$ should be higher than the eigenvalue of the kinetic energy operator. Secondly, we compare our calculations with the energy $V_c(\nu)$ calculated by Baranov *et al.* [25]. We find that the DDI energies of the WC with $\alpha=1/4$ and $d=0$ are $V_c(\nu=1/3) = 0.5033(1/3)^{3/2} D/\lambda^3$ and $V_c(\nu=1/5) = 0.3590(1/5)^{3/2} D/\lambda^3$, respectively. Our calculations are consistent with the results from Baranov *et al.* in the small filling factor regime. As the filling factor is increasing, our calculations start deviating from the results from Baranov *et al.*, which is valid in the small filling-factor regime. Thirdly, as the lowest Landau level completely filled, i.e., $\nu=1$, the DDI energies of the dipolar gas and WC should be equal. We got $V_{gas}(\nu=1)=0.6266 D/\lambda^3$ and $V_c(\nu=1)=0.6266 D/\lambda^3$ as $d=0$. From the above comparisons we can see that Löwdin's formula can give us accurate DDI energies of the WC.

To find the phase boundary between the FQH liquid and the WC, we have to calculate the total energy from the FQH liquid. Note that the lowering in energy of the liquid from LLM is substantially smaller than that in solid [27]. No LLM effect is considered in finding the ground-state energy of the FQH state. In the lowest Landau level, the kinetic energy per particle, $K_\ell$, of the FQH state is $\hbar\omega_c/2$. The DDI energy per particle, $V_\ell(\nu)$, of the FQH liquid is $V_\ell(\nu) = (\nu/2)\int_0^\infty rg(r)U(r)drdt$, where $g(r)$ is the radial function of the Laughlin liquid state which has been computed by the Monte Carlo method. $g(r)$ can be approximated as [31]:

$$g(\mathbf{r}) = 1 - e^{-r^2/2} + \sum_{m=1}^{\infty} \frac{C_m}{m!}[1-\exp(-im\pi)]\left(r^2/4\right)^m e^{-r^2/4}, \quad (6)$$

where the expansion coefficients $C_m$, obtained by fitting the Monte Carlo data, are listed in reference 31. The DDI energies of the FQH liquid with $d/\lambda=0$ are $V_\ell(\nu=1/3) = 0.3676(1/3)^{3/2} D/\lambda^3$ and $V_\ell(\nu=1/5)=0.3354(1/5)^{3/2} D/\lambda^3$, respectively. These energies are in excellent agreement with the energies from Baranov *et al*. [25]. By comparing $V_c(\nu)$ with $V_\ell(\nu)$ for cases $\nu=1/3$ and $\nu=1/5$, the FQH liquid is indeed lower in energy than the WC in the lowest-Landau-level approximation, which is consistent with the conclusion from Baranov *et al*. [25].

Without considering the LLM, we have shown that the stable phase of rotating dipolar fermions characterized by $\nu=1/3$ or $\nu=1/5$ is the FQH liquid. In order to compare variational $V_c(\nu)$ with $V_\ell(\nu)$ in the regime that the effect of LLM is important, the energy and length scales given by the cyclotron energy $\hbar\omega_c$ and the magnetic length $\lambda$ should be expressed in terms of effective atomic units. The dimensionless ground-state energy $\Omega$ of any state is defined by subtracting a constant energy $\hbar\omega_c/2$ to the total energy $E$ and given by:

$$\Omega = \frac{E - 0.5\hbar\omega_c}{E_B} = (\frac{K}{\hbar\omega_c} - 0.5)\left(\frac{2\pi r_s}{\nu}\right) + r_s^3 \left(\frac{2\pi}{\nu}\right)^{3/2} \left(\frac{V}{D/\lambda^3}\right), \quad (7)$$

where $E_B = D/a_B^3$; $K$ and $V$ are the kinetic energy and DDI energy per particle of the system, respectively. $K = K_c$ and $V = V_c(\nu)$ ($K = K_\ell = \hbar\omega_c/2$ and $V = V_\ell(\nu)$) for the Wigner crystal (for the FQH state). From Eq. (7), we can see that the DDI energy dominates the kinetic energy and the system will crystallize into a lattice [26] as $r_s$ becomes bigger. In Fig. 1, we show the vaiational ground-state energy $\Omega$ of the WC as a function of the variational parameter $\alpha$ for $r_s=4$ and $r_s=7$ at $\nu=1/5$ and the thickness $d/\lambda = 0$. We have also shown the corresponding ground-state energy of the FQH liquid in Fig. 1 (see the solid horizontal lines). We can optimize the ground-state energy of the WC from the variation of $\alpha$ parameter. For $r_s=4$ (or $r_s=7$) the FQH-state energy is lower (or higher) than the optimized energy of the WC. The variational ground-state energies of the WC for $r_s=4$ and $r_s=5$ at $\nu=1/3$ and $d=0$ has the same qualitative behavior as $\nu=1/5$ (see Fig. 2). The effect of LLM on lowering the ground-state energy of the WC is increasing when $r_s$ is increasing. Above some critical dimensionless length $r_s^c$, $r_s^c=5.108$ for $\nu=1/5$ and $r_s^c=4.642$ for $\nu=1/3$, the WC becomes the stable phase of the system if the thickness $d=0$. The critical dimensionless length $r_s^c$ as a function of the thickness is shown in Fig. 3. The critical dimensionless length $r_s^c$ is increasing as the thickness of dipolar gases is increasing. This behavior implies that the WC with a larger thickness is less stable than the WC with a smaller thickness. The crystal and liquid phases are stable in the larger and smaller $r_s$ regime, i.e., the higher and lower density regime, respectively. For rotating dipolar gases the crystal phase is expected at *high density*, which is consistent with the argument that the non-rotating dipolar gases with a *high density* have a crystal order [11, 12]. This result is inconsistent with the result from Baranov *et al*. [25], where the effect of LLM on the WC is ignored. We believe that the effect of LLM is important in the high density regime of rotating dipolar fermions.

In conclusion, we have studied the possible phases of rotating dipolar Fermi gas in the FQH regime. Two kinds of states, the FQH liquid and WC, are used to determine the ground-state energy of the stable phase in the FQH regime. Contrary to the lowest Landau approximation, where the effect of LLM is ignored, we find that the LLM is an important effect on determining the stable phase in the FQH regime. The LLM is a strong/weak effect in the high/low density regimes, respectively. We apply a variational ansatz, which includes the effect of LLM, to optimize the ground-state energy of the WC. We show that the LLM will lower the energy of the WC significantly in the high-density regime. This lowering in energy of the WC makes the WC becoming the stable phase in the high density regime of the FQH effect, whereas the stable phase is the FQH liquid in the low density regime. Therefore, the crystal phase is expected at *high density* for rotating dipolar gases, which is consistent with non-rotating dipolar gases, but is inconsistent with the low-density conclusion where the effect of LLM is ignored. We conclude that the effect of LLM should be considered in finding the stable phase of rotating dipolar Fermi gas in the FQH regime.

We acknowledge the financial support from the National Science Council (NSC) of Republic of China under Contract No. NSC96-2112-M-034-002-MY3. S. C. thanks the support of the National Center for Theoretical Sciences of Taiwan during visiting the center.

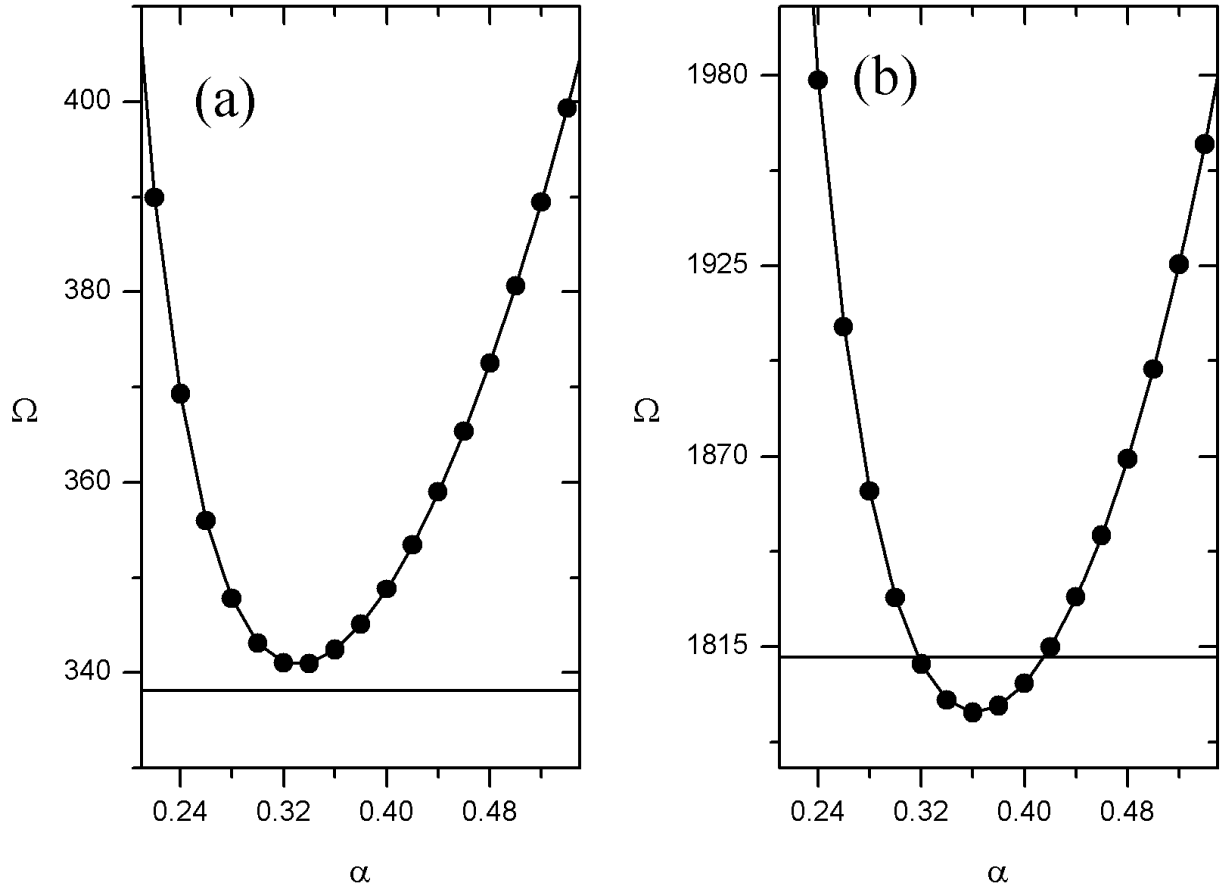

Fig. 1. Dimensionless Wigner-solid energies vs. variational parameter $\alpha$ for $\nu=1/5$ at (a) $r_s=4$, (b) $r_s=7$. The sample thickness $d=0$. Black dots are variational energies of the Wigner crystal. The solid horizontal line is the ground-state energy of the quantum Hall liquid.

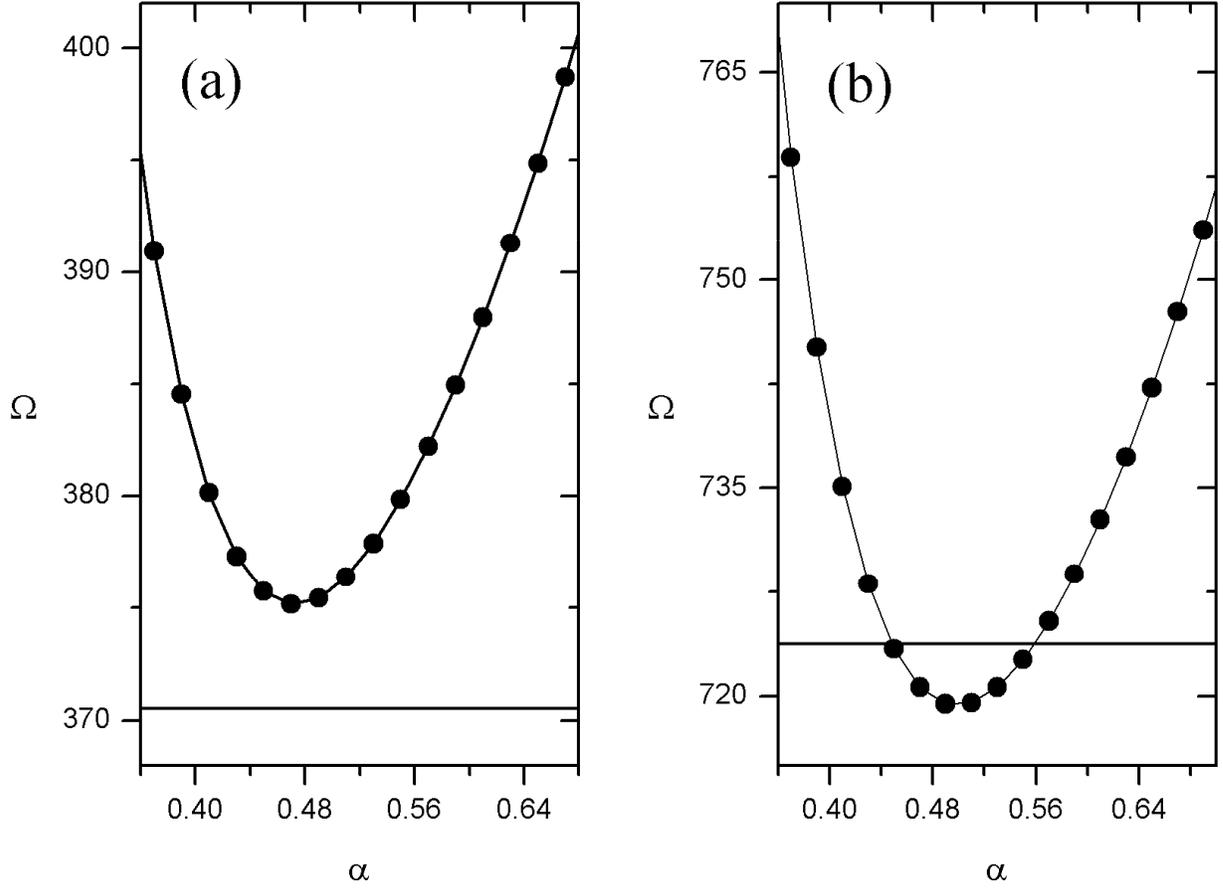

Fig. 2. Dimensionless Wigner-solid energies vs. variational parameter $\alpha$ for $\nu=1/3$ at (a) $r_s=4$, (b) $r_s=5$. The sample thickness $d=0$. Black dots are variational energies of the Wigner crystal. The solid horizontal line is the ground-state energy of the quantum Hall liquid.

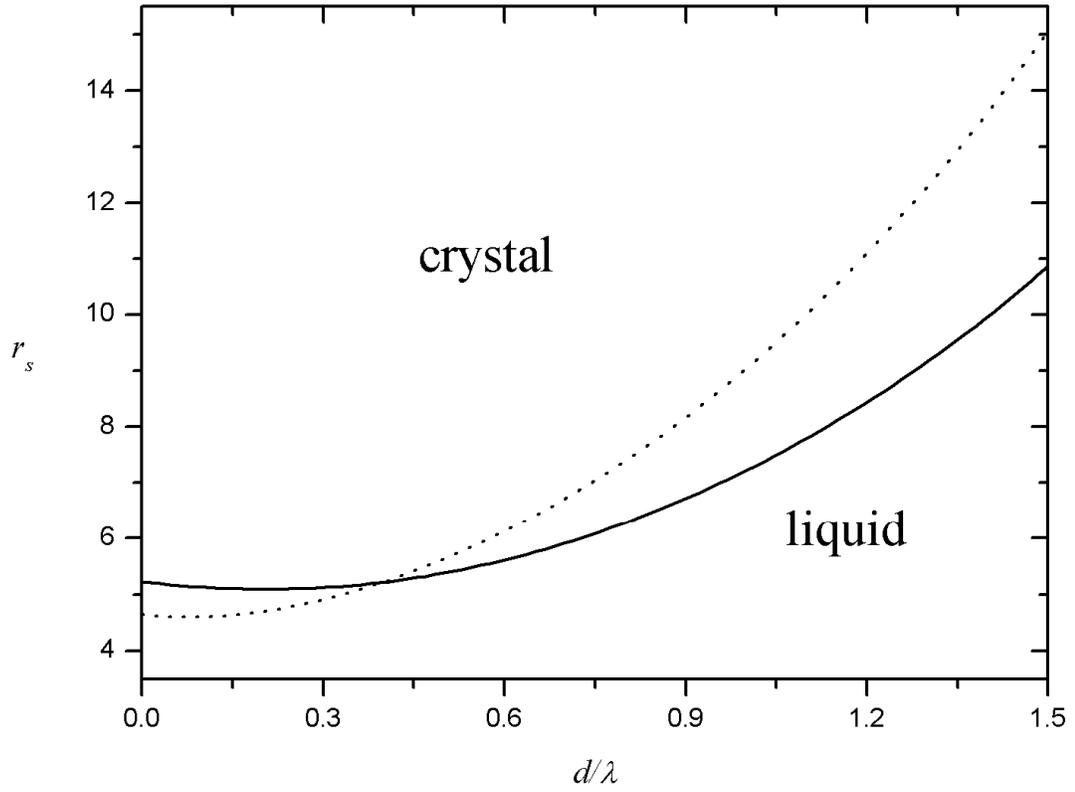

Fig. 3. Phase diagram of rotating dipolar gases. The solid and dotted lines indicate the critical dimensionless length $r_s$ as a function of the thickness for $\nu=1/5$ and $\nu=1/3$, respectively.